\documentclass[prl,twocolumn,showpacs,nofootinbib]{revtex4}
\usepackage{graphicx}
\usepackage{epstopdf}
\usepackage{epsfig}
\usepackage{color}

\newcommand{\be}{\begin{equation}}
\newcommand{\ee}{\end{equation}}
\renewcommand{\r}{{\bf r}}

\begin{document}


\title{Cooperative effects and photon localization  in atomic gases: \\ The two-dimensional case}
\author{A. Gero$^{1,2}$ and E. Akkermans$^{1}$ }
\affiliation{$^{1}$Department of Physics, Technion - Israel Institute of Technology, 32000 Haifa, Israel \\ $^{2}$Department of Education in Technology and Science, Technion - Israel Institute of Technology, 32000 Haifa, Israel}

\begin{abstract}
We study photon escape rates from two-dimensional atomic gases while taking into account cooperative effects between the scatterers. Based on the spectrum of the random Euclidean matrix $U_{ij}=J_0(x_{ij})$, where  $x_{ij}$ is the dimensionless random distance between any two atoms,  a scaling behavior for the escape rates is obtained.  This behavior is  similar to the one obtained in  the three-dimensional case, but qualitatively different from the result obtained in  one dimension.

\end{abstract}

\pacs{42.25.Dd, 42.50.Nn, 72.15.Rn}

\date{\today}

\maketitle

\section{I. Introduction}
Superradiance and subradiance, first described by Dicke \cite{dicke} in 1954, stem from multiple exchanges of  a photon between atoms. These cooperative effects change significantly the lifetime and the radiative level shift of the atoms \cite{stephen,lehmberg, gross}. Over the years, superradiance has been investigated in the context of  Bose-Einstein condensate \cite{baumann, kampel}, molecules \cite{lin}, cold atoms \cite{PRL1,PRA, bux,bienaime} and nuclei \cite{Roehlsberger}.

 Recently,  the  contribution of cooperative effects to  photon localization, which shows up as an overall decrease of photon escape rates from a random medium, has been studied \cite{PRL2, EPL}. It has been shown that for a three-dimensional atomic gas, photon localization, i.e., the trapping of photons in the gas, occurs as a smooth crossover between two limits: the single atom limit where the photons are delocalized in the atomic ensemble and spontaneous emission of independent atoms occurs, and  the opposite limit where the photons are localized and trapped  in the gas for a very long period of time. These two limits are connected by a crossover rather than  a disorder-driven phase transition as expected on the basis of the theory of Anderson localization \cite{PRL2}. Moreover,  for a one-dimensional disordered atomic system, due to cooperative effects and not disorder,  the  single-atom limit is never reached and the photons are always localized in the gas \cite{EPL}. 

In this paper we study photon escape rates from a two-dimensional atomic gas and compare them to those obtained in one and three dimensions.  To that purpose, we follow the method, based on the  Marchenko-Pastur law \cite{MPL}, that has been developed   for diagonalizing Euclidean random matrices in three dimensions \cite{skipetrov}. We will  prove that not only this method  is applicable  for the two-dimensional case, but it yields even better approximations compared to those obtained in  three dimensions. We will show that the two-dimensional case is  similar to  the three-dimensional case, but qualitatively different from the one-dimensional case. We will also show that the function which measures photon localization in one and three dimensions  exhibits a scaling behavior in the two-dimensional case as well.

The paper is organized as follows: We start, in Section II, by describing the model which consists of identical atoms placed at random positions in a  radiation field. Then, in Section III, we briefly review cooperative effects for a pair of atoms.  In Section IV, we  go beyond the case of two atoms and introduce the method for calculating  photon escape rates from atomic gases. Next, in Section V we apply the method to the two-dimensional case and  compare, in Section VI, the obtained results to those of  one  and three dimensions. Finally our findings are summarized in Section VII.

\section{II. Model}
We consider an ensemble of $N\gg1$ identical atoms, placed at random positions $\textbf{r}_{i}$ and coupled to the  radiation field.
Atoms are taken  as non-degenerate, two-level systems denoted by $|e\rangle$ for  the excited state and $|g\rangle$ for the ground state. The energy
separation between the two levels, including radiative shift,  is
$\hbar\omega_{0}$ and the natural width of the excited level is
$\hbar \Gamma_{0}$. 

The interaction 
between the radiation field and the electric dipole moments of the atoms is
\be V=-\sum_{i=1}^{N}
{\textbf{d}}_{i}\cdot \textbf{E}(\textbf{r}_{i})\label{eqc2}, \ee where
${\textbf{E}}(\r )$ is the electric field operator
and
${\textbf{d}}_{i}=e\textbf{r}_i$ is the electric dipole moment operator of the
$i$-th atom.

We assume that the typical speed of the atoms  is small compared to
$\Gamma_{0}/k$ but large compared to $\hbar k / \mu$, where $k$ is the radiation wave number and $\mu$ is
the mass of the atom, so that it is possible to neglect the
Doppler shift and recoil effects. In addition, retardation effects are neglected, thus each atom can influence the others instantaneously.

\section{III. Case of  two atoms}
The absorption of a photon by a pair of de-excited atoms, located at $\r_{1}$ and $\r_{2}$,  leads to a configuration where the two atoms, one excited
and the second in its ground state, undergo multiple exchanges of a
photon, giving rise to an effective interaction potential and  a
modified lifetime as compared to independent atoms \cite{stephen,lehmberg}.
This process is described  using the basis of the Dicke states
\cite{dicke}, where the singlet Dicke state is \be|00
\rangle = {1 \over \sqrt{2}} [ |e_{1} g_2 \rangle - |g_{1} e_{2}
\rangle ] \label{eq8a}\ee and the triplet Dicke states are
\begin{eqnarray}
|11 \rangle &=& |e_{1} e_2  \rangle,  \nonumber \\ |10 \rangle &=& {1 \over \sqrt{2}} [ |e_{1} g_2 \rangle + |g_{1} e_{2} \rangle], \nonumber  \\ |1-1\rangle &=&|g_{1} g_2 \rangle.
\label{eq8b}
\end{eqnarray}
The singlet state $|-\rangle=|00 \rangle$ and the
triplet state $|+\rangle=|10 \rangle$ both correspond to one atom in the
excited state and the other in the ground state, but $|00 \rangle$
is anti-symmetric  where $|10 \rangle$ is symmetric under an
exchange of the atoms. 

In a three-dimensional system, the expression for the cooperative spontaneous emission rate or the inverse lifetime, $\Gamma^{\pm}$, of these states is \cite{stephen,lehmberg}
\be {\Gamma^\pm \over \Gamma_{0}}
= 1\mp {3 \over 2}  \left[ - p {\sin k_0 r \over k_0 r} + q
\left( {\sin k_0 r \over (k_0 r)^3} -  { \cos k_0 r \over (k_0
r)^2 } \right) \right] \label{eqa27},\ee
where $k_{0}=\omega_{0}/c$, $\r=\r_{1}-\r_{2}=(r,\theta,\varphi)$, $ p(\theta) = \sin^2\theta$ and $q(\theta) = 1-3\cos^2\theta.$
 Averaging (\ref{eqa27}) upon the random orientations of the pair of atoms \cite{PRL1,PRA}, yields
\be {\Gamma^\pm \over \Gamma_{0}}= 1\pm \frac{\sin k_0r}{k_0 r},\label{eqaa37}\ee namely,  the cooperative spontaneous emission rate in the case where the atoms are coupled to a scalar radiation field \cite{vries}. 
As the scalar model has the advantage of being easier to handle, from now on we will consider the diploar interaction of atoms and  a scalar radiation field.

For the two-dimensional case, considered here, we assume that the atoms are constrained within a two-dimensional square, so that ${\textbf{d}}_{i}$ is a two-dimensional vector. Additionally, the photon exchanges  between the atoms are restricted to occur in the plane of the square. Under these constraints the cooperative spontaneous emission rate is \cite{book}
\be {\Gamma^\pm \over \Gamma_{0}}
= 1\pm J_0(k_0 r) \label{SF22},\ee
where $J_0(x)$ is the Bessel function of the first kind.
For comparison, for the one-dimensional case  \cite{milonni2} \be {\Gamma^\pm \over \Gamma_{0}}
= 1\pm \cos k_{0}r \label{eqa37}.\ee


 When the atoms are close enough $(k_{0}r\ll1)$ the Dicke limit, namely $\Gamma^{\pm}=(1\pm1)\Gamma_{0}$, is reached in all cases. On the opposite limit, when the atoms are well separated, the single-atom emission rate is recovered from both  (\ref{eqaa37})-(\ref{SF22}) for any value of $k_{0}r\gg1$. In the one-dimensional case, however, the single-atom emission rate cannot be obtained for an arbitrary fixed value of $k_{0}r\gg1$, but by averaging (\ref{eqa37}) over $k_{0}r\gg1$.  This fundamental difference will be reflected  in the comparison of photon escape rates from  atomic gases in different dimensions, carried out later on.

\section{IV. Photon escape rates from atomic gases}

\subsection{A. Photon escape rates}

Cooperative spontaneous emission rates of more than two atoms, namely photon escape rates from an atomic gas with $N \gg 1$ atoms, with a single excitation, can be determined  by the time evolution of the ground state population  associated with the reduced atomic density operator $\rho$ of the gas \cite{tallet1, tallet2, PRL2, EPL}: 
\be \frac{d\langle G| \rho |G \rangle}{dt}= \Gamma_{0}\sum_{ij}U_{ij}\langle G|\Delta_{j}^{-}\rho \Delta_{i}^{+}|G\rangle \label{evo},\ee
where $|G\rangle=|g_{1}, g_{2},...,g_{N}\rangle$, $\Delta_{i}^{+}=(|e\rangle \langle g|)_i$ is the raising operator of atom $i$ and $\Delta_{i}^{-}=(|g\rangle\langle e|)_i$ is the corresponding lowering operator.
For the two-dimensional case \be U_{ij}=J_0( k_0 r_{ij}),\label {UU}\ee while for the three-dimensional gas
 \be U_{ij}=\frac{\sin k_0r_{ij}}{k_0r_{ij}},\label {u}\ee  and for the one-dimensional case  \be U_{ij}=\cos k_0 r_{ij},\label {uu} \ee where $r_{ij}=|\r_{i}-\r_{j}|$ is the random distance between any two atoms. 

The eigenvalue equation of $U_{ij}$ is
\be \sum_{j=1}^{N}U_{ij}u_{j}^{(n)}=\Gamma_{n}u_{i}^{(n)}\label{eqc23},\ee where $\Gamma_{n}$ is the $n$-th dimensionless  eigenvalue associated with $u^{(n)}$, the $n$-th eigenfunction.
Using orthonormality of the eigenfunctions, we can rewrite (\ref{evo}) as
\be \frac{d\langle G| \rho |G \rangle}{dt}= \Gamma_{0}\sum_{n=1}^{N}\Gamma_{n}\langle G|\Delta_n^{-} \rho \Delta_n^{+}|G\rangle\label{eqc25},\ee where the  collective raising and lowering operators are
$\Delta_{n}^{\pm}=\sum_{i=1}^{N}u_{i}^{(n)}\Delta_{i}^{\pm}$.
Thus,  the eigenvalues of the coupling matrix $U$ can be interpreted as the photon escape rates from the atomic gas \cite{tallet1, tallet2, ernst1, ernst2}.


\subsection{B. Average density of photon escape rates}

The elements of the matrix $U$  are function of the distance between  
random points in an Euclidean space. This type of matrices has been referred to Euclidean random matrices, and it has been studied thoroughly \cite{mezard, amir, kanter}.
The average density of eigenvalues of $U$  is 
\be P(\Gamma)=\frac{1}{N}\overline{\sum_{n=1}^{N}\delta(\Gamma-\Gamma_{n})}\label{eqc30},\ee where the average, denoted by $\overline{\cdot
\cdot \cdot}$, is taken over the spatial configurations of the atoms.
The density, \be P(\Gamma)=-\frac{1}{\pi}\mbox{Im}G_U(z=\Gamma+i\epsilon)\label{eqc31},\ee can be obtained from the trace of the resolvent matrix \be G_U(z)=\frac{1}{N}\overline{\mbox{Tr}(zI-U)^{-1}},\label{eqc311}\ee where $I$ is the $N\times N$ unit matrix.

 Let us consider the two limits:
\begin{enumerate}
\item The Dicke limit ($k_{0}r_{ij}\ll1$)
 
Here $U$, given either by (\ref{UU}),(\ref{u}) or (\ref{uu}), reduces to \be
U = \left(\begin{array}{llllllll} 1 & 1 & \cdots & 1 \\
1 & 1 & \cdots & 1 \\
\vdots & \vdots &  & \vdots \\
1 & 1 & \cdots & 1
\end{array}\right) \
\label{matrice}
\ee
 and \be G_U(z)=\frac{1}{N}\left(\frac{1}{z-N}+\frac{N-1}{z}\right),\label{G}\ee thus \be P(\Gamma)=\frac{1}{N}[\delta(\Gamma-N)+(N-1)\delta(\Gamma)]\label{eqc33}.\ee
In this case the eigenvalue $\Gamma =0$ is the $(N-1)$-degenerate subradiant mode and $\Gamma = N$ is the non-degenerate superradiant mode.

\item Dilute gas ($k_{0}r_{ij}\gg1$)

For the two- and three-dimensional cases, $U=I$ and $G_U(z)=(z-1)^{-1}$, thus \be P(\Gamma)=\delta(\Gamma-1).\label{eqc32}\ee
In this limit the single-atom spontaneous emission rate is recovered. For one dimension, however,  $U\neq I$ and the single-atom limit is never reached, as noted in Section III.

\end{enumerate}

\subsection{C. Measure of photon localization}
In order to characterize $P(\Gamma)$  and obtain a measure of photon localization we use  the  following  function \be C=1-2\int_1^\infty d\Gamma P(\Gamma), \label{C}\ee normalized to unity. $C$  measures the relative number of states having a vanishing escape rate.  The function $C$ exhibits a scaling behavior over a broad range of system size and density both in  one \cite{EPL} and three dimensions  \cite{PRL2}. Later on, we will show that in the two-dimensional case $C$ exhibits a scaling behavior as well.  

In the Dicke limit ($k_{0}r_{ij}\ll1$), it is straightforward from (\ref{eqc33})  that $C=1-2/N$ in all dimensions. Thus, for $N\gg1$, $C=1$ and the photon is localized in the gas. In the opposite limit, obtained only in the two- and three-dimensional cases, (\ref{eqc32}) yields $C=0$, indicating delocalization \cite{note}.

 It should be noted that  the mean value of $\Gamma$ hardly  characterizes $P(\Gamma)$  since $\Gamma_{mean}=[\mbox{Tr}(U)]/N$ and $\mbox{Tr}(U)=N$ in all dimensions, therefore  $\Gamma_{mean}=1$ regardless of the system parameters. In the next section we will show that $C$, in a certain limit, is a function of the variance of $P(\Gamma)$.

\subsection{D. Decomposition of  $U$ and  Marchenko-Pastur distribution}
 Beyond the Dicke and the dilute gas limits, it is difficult to obtain a non-approximate analytic expression of the distribution $P(\Gamma)$, except for the case of one dimension \cite{EPL}. There,  the atoms are distributed along a line, and the coupling matrix (\ref{uu}) can be rewritten as
$U=\frac{1}{2} H^{\dag}H$, where $H$ is the $2\times N$ matrix defined by
$ H_{0j}=e^{ik_0r_j}$ and
$H_{1j}=e^{-ik_0r_j}$. As $U$ is a real symmetric matrix,  its spectrum is given exactly by the spectrum of the $2 \times 2$ matrix $U^{\dag}$ plus $N-2$ vanishing eigenvalues. 

In higher dimensions, a useful approach consists in looking for the $N \times N$ matrix $U$ under the form of a product 
\be
U = H \, T \, H^\dagger,
\label{decomposition}
\ee
where $T$ is a $M \times M$ matrix while $H$ is a $N \times M$ rectangular matrix. This approach, promoted  for the three-dimensional case \cite{skipetrov}, is of wide application in random matrix theory methods applied to wireless communication \cite{couillet}. The product form for $U$ results from a decomposition where the randomness is contained in the  matrix $H$, while the square and non-random  matrix $T$ counts the $M$ transverse modes of the $d$-dimensional cavity which contains the atoms, assuming periodic boundary conditions. Important results have been obtained for the asymptotic distribution of random matrices of the form (\ref{decomposition}) in the context of the free probability theory \cite{voiculescu, tulino,couillet}.

We wish to obtain the distribution $P( \Gamma)$ of the matrix $U$ in the limit $(N,M) \rightarrow \infty$ so that the ratio $ N / M$ is constant. Beforehand, we introduce some definitions and notations. We define, for any square random matrix $U$,  the quantity $G^{-1} _U (z)$ as the functional inverse (with respect to the composition of functions) of  the resolvent $G_U(z)$, namely,  $ G^{-1} _U \left[ G_U (z) \right] = z$. We then define  the $R$-transform, also called self-energy, by 
\be
R_U (z) = G^{-1} _U (z) - {1 \over z},
\label{R}
\ee
and the $S$-transform, $S_U (z)$, by means of the implicit relation
\be
G_U \left( {z +1 \over z S_U (z) } \right) = z S_U (z).
\label{S}
\ee

We now give two important results which will prove useful \cite{voiculescu, tulino, couillet}. The first states that for asymptotically free matrices  $A$ and  $B$, their $S$-transforms fulfill 
\be
S_{AB} (z) = S_A (z) \, S_B (z).
\label{productS}
\ee
This shows the interesting  feature of the $S$-transform, which allows to disentangle the product of matrices. Moreover, for any $N \times M$ matrix $A$ and $M \times N$ matrix $B$ with $(N,M) \rightarrow \infty$ and  constant $\beta= N/M$, one has \cite{tulino}
\be
S_{AB} (z) = {\beta (z + 1) \over 1 + \beta z} \, S_{BA} (\beta z).
\label{product}
\ee

The second important result states that  a random $N\times M$ matrix $H$ whose entries are zero-mean $\it i.i.d$ (independent and identically distributed) random variables with variance $1/N$ is such that in the limit $(N,M) \rightarrow \infty$ and constant $\beta= N/M$, the distribution of eigenvalues of the square $M \times M$ product matrix $H^\dagger \, H$ converges almost surely to the Marchenko-Pastur law \cite{MPL,tulino},
\be
P^{MP} (\Gamma) = \left( 1 - \beta \right)^+ \, \delta (\Gamma) + {\beta \, \sqrt{\left( \Gamma- a \right)^+ \, \left( b-\Gamma \right)^+ } \over 2 \pi \, \Gamma},
\label{MP}
\ee
with $x^+ = \mbox{max} (0,x)$ and $(b,a) = \left( 1 \pm 1/\sqrt{\beta} \right)^2$. This result is  analogous  to the central limit theorem. 
Finally, the $R$ and $S$ transforms of the Marchenko-Pastur law are given, respectively,  by \cite{tulino}
\be
R^{MP} (z) = {\beta \over \beta -  \, z }
\label{RMP} 
\ee
and
\be
S^{MP} (z) = {\beta \over \beta +  \, z }. 
\label{SMP}
\ee

We now apply these results to obtain an important relation between $R_U(z)$ and $G_T(z)$,  where the matrix $U$ is given by the decomposition (\ref{decomposition}). Using (\ref{productS})-(\ref{product}) we have
\be 
S_U (z) = {\beta (z + 1) \over 1 +\beta z} \,  S_{H^\dagger H} \left( \beta z \right) \, S_T \left( \beta z \right),
\label{SU1}
\ee
and with the help of  (\ref{SMP})
\be
S_U (z) = {\beta \over 1 + \beta z} \,  S_T \left( \beta  z \right).
\label{SU2}
\ee
Next, we apply (\ref{S}) to the matrix $T$ and the variable $\beta z$, namely,
\be
G_T \left( {1 +\beta z \over \beta z S_T ( \beta z) } \right) = { \beta z} S_T (\beta z)
\label{Szbeta}.
\ee
Defining a new variable, 
\be
{ 1 \over Z} = {1 +\beta z\over \beta z S_T (\beta z) } ,
\label{newvariable}
\ee
we obtain from (\ref{SU2}) that 
\be
{ 1 \over Z} = {1 \over z \, S_U (z)} \label{ZZZ},
\ee
and from (\ref{Szbeta})-(\ref{newvariable})  
\be
G_T \left( {1 \over Z} \right) = (1 + \beta z) \, Z.
\label{GTZ}
\ee
We use  (\ref{S}) again,  this time for the matrix $U$, and with the help of (\ref{ZZZ}) we obtain,
\be
z = Z \, G^{-1} _U (Z) -1.
\ee
Inserting the last expression into (\ref{GTZ}) and solving for $G^{-1} _U (Z) - 1/Z$ yield (after a change of variables)
\be
R_U (z) = {1 \over \beta z} \, \left[ {1 \over z} \, G_T \left( {1 \over z} \right) -1\right]. 
\label{GU-1}
\ee
This relation between the resolvent of $T$ and the self-energy of $U$ will be used in the next section in order to obtain the eigenvalue distribution $P(\Gamma)$ of the matrix $U$ in the two-dimensional geometry.

\section{V. Two-dimensional geometry}
 Now, we  investigate the distribution of the eigenvalues  of (\ref{UU}) and the corresponding function $C$, both numerically and analytically. 

\subsection{A. Spectrum of $U_{ij}=J_0(k_0r_{ij})$}

 We consider $N\gg1$ atoms enclosed in a square $L^{2} \equiv(2\pi a /k_0)^{2}$, thus defining the dimensionless length $a$. The atoms are randomly distributed with a uniform density $n=N/{L}^{2}$ and the corresponding coupling matrix is given by (\ref{UU}). The average density of photon escape rates, obtained for many random configurations of the atoms, is presented in Fig. 1 for different values of the dimensionless density $W=N/2\pi a^2$ and $a\geq1$. The Dicke limit ($a\ll1$) is shown in Fig. 2 and the corresponding $P(\Gamma)$  is given by (\ref{eqc33}).

\begin{figure}[ht]
\centerline{ \epsfxsize 9cm \epsffile{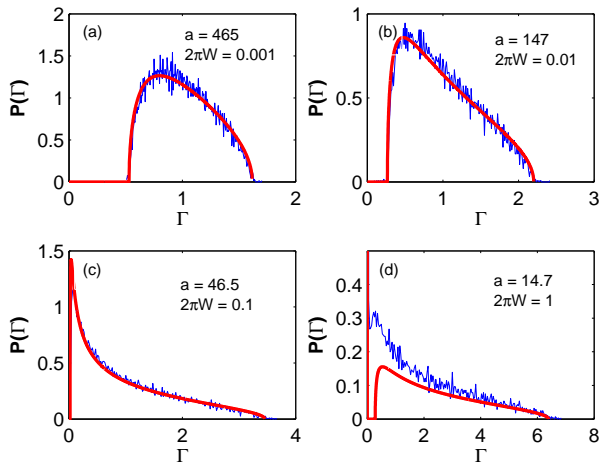} } \caption{\em (Color
   online)Behavior of $P (\Gamma)$ for different values of  $W$, $a=M/2\pi \geq 1$ (see text for a definition of M) and for $N=216$ in a two-dimensional geometry. The solid line is calculated using the Marchenko-Pastur law (\ref{PPP}).}
 \label{fig1}
\end{figure}

\begin{figure}[ht]
\centerline{ \epsfxsize 9cm \epsffile{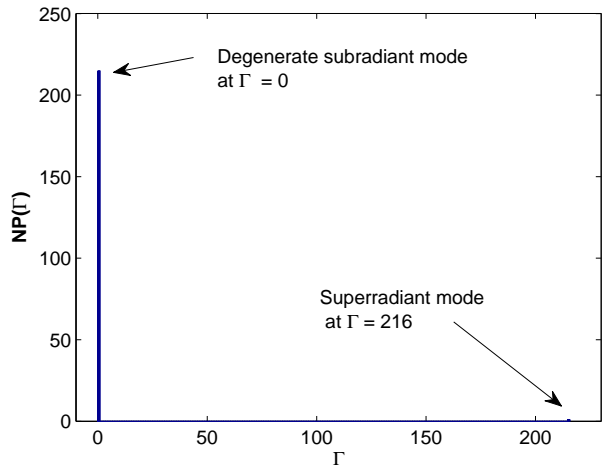} } \caption{\em (Color
   online)Behavior of $NP (\Gamma)$ in the Dicke limit ($a=0.015$) for $N=216$ in a two-dimensional geometry. The distribution is described by (\ref{eqc33}). }
 \label{fig2}
\end{figure}

In order to obtain the eigenvalue distribution of  $U$  away from the Dicke limit, according to relation (\ref{GU-1}), we
should  calculate $G_T(z)$, the resolvent of the matrix $T$  introduced in (\ref{decomposition}). To that purpose, we follow the approach of  ref. \cite{skipetrov} and choose \be H_{im}=\frac{1} {\sqrt{N}}  e^{i\textbf{q}_m\cdot \textbf{r}_i}, \ee where  $\textbf{q}_m=\{ q_{mx}, q_{my}\}$ with $q_{mi}=m_i\frac{2\pi}{L}$ and $m_i=\pm 1, \pm2, ...$. Thus, $T$ is an approximate representation of the two-dimensional Fourier transform of $U$. It should be noted that the elements of the  random matrix $H$ obey the conditions necessary to obtain (\ref{MP}).

The two-dimensional Fourier transform of $U_{ij}=J_0(k_0r_{ij})$ in the square $S=L^2$ is
\be T(\textbf{q}_m, \textbf{q}_n)=\frac{N}{ S^2} \int d^2\textbf{r}_i\int d^2\textbf{r}_j\ J_0(k_0r_{ij}) \ e^{-i\textbf{q}_m\cdot \textbf{r}_i+i\textbf{q}_n\cdot \textbf{r}_j}.\ee
 By changing to new integration variables $\textbf{R}=(\textbf{r}_i+\textbf{r}_j)/2$ and $\textbf{r}=\textbf{r}_j-\textbf{r}_i$ and confining  the integration over $\r$ to $|\r|<L/2\alpha$, where $\alpha
$ a numerical constant to be determined later on, we obtain the following approximation
\begin{eqnarray}
 T(\textbf{q}_m, \textbf{q}_n)\simeq\frac{N}{ S^2} \int d^2\textbf{R}\ e^{-i(\textbf{q}_m-\textbf{q}_n)\cdot \textbf{R}} \nonumber \\
 \times  \int d^2\textbf{r}\ J_0(k_0r)\ e^{i(\textbf{q}_m+\textbf{q}_n)\cdot \textbf{r}/2}. 
\end{eqnarray}

In order to calculate the second integral we use that the two-dimensional Fourier transform of a circularly symmetric function $f(r)$  is equivalent, up to $2\pi$, to the Hankel transform of order zero, namely $\mbox{H}_0[f(r)]=\int_0^\infty dr \  rf(r)J_0(qr)$. Since $\mbox{H}_0[J_0(k_0r)]=\delta (q-k_0)/k_0$,  the value of the second integral in an infinite square is  $2\pi\delta (q_m-k_0)/k_0$.  In our case, however, the integration over $r$  is limited to $|\r|<L/2\alpha$. Thus, we approximate the delta function by a sinc function and have
\be T(\textbf{q}_m, \textbf{q}_n)\simeq\frac{N}{ S}\delta(\textbf{q}_m- \textbf{q}_n) \frac{\pi L}{\alpha k_0} \ \mbox{sinc} \left[(q_m-k_0)\frac{L}{2\alpha}\right]. \label{t}\ee
For comparison, the calculation  of  $T$ in the  three-dimensional case \cite{skipetrov}, where $U$ is given by (\ref{u}), yields an additional sinc function, peaked around $q_m=-k_0$. This additional term has been  discarded by the authors of  \cite{skipetrov} in their calculation. Here, there is no such approximation, hence the results in the two-dimensional case are expected to be more accurate compared to those in three dimensions.

The sinc function in (\ref{t}), peaked around $q_m=k_0$ for $L\gg1$,  limits the number of  $\textbf{q}_m$'s that contribute to $T$. Therefore, we take into account  only the contribution of $\textbf{q}_m$'s within a two-dimensional spherical shell of radius $k_0$ and thickness $2 \alpha / L$, namely  $M=\alpha k_0L/ \pi$  modes (if $M$ is not an integer, we refer to the integer part of $M$). Furthermore, for all of these  $\textbf{q}_m$'s we approximate (\ref{t}) as


\be T(\textbf{q}_m, \textbf{q}_n)\simeq \frac {N}{M}\delta(\textbf{q}_m- \textbf{q}_n).\label{tt}\ee
It should be noted that in the three-dimensional case \cite{skipetrov}, $M$ varies like $( k_0L)^2$. This difference  is significant as it causes localization to occur earlier in  the two-dimensional case compared  to the three-dimensional case, as we will see in Section VI.  

Next, we consider in (\ref{tt}) only the modes $\textbf{q}_m= \textbf{q}_n$ for which $m=n$. In this case, $T$ can be represented by the following $M\times M$ matrix,
\be T_{mn}\simeq \frac {N}{M}\delta_{mn}.\label{Tmatrix} \ee
The  resolvent (\ref{eqc311}) of $T$ is thus
given by
\be G_T(z)=\left(z-\frac{N}{M}\right)^{-1}.\label{GT}\ee
Substituting (\ref{GT}) in relation (\ref{GU-1}) yields the self-energy of $U$,
\be R_U(z)=\left(1-\frac{N}{M}z\right)^{-1}.\ee
The last result is the self-energy of the Marchenko-Pastur law (\ref{RMP}) with $\beta=M/N$. Therefore, the  spectrum of $U_{ij}=J_0(k_0r_{ij})$ is approximated for $M\gg1$ by
\be P(\Gamma)\simeq\left(1-\frac{M}{N}\right)^+ \delta(\Gamma)+ \frac{\sqrt{(\Gamma-\Gamma_-)^+(\Gamma_+-\Gamma)^+}}{2\pi \frac{N}{M}\Gamma},\label{PPP}\ee where $\Gamma_\pm =(1 \pm \sqrt{N/M})^2$, as shown in Fig. 1.
Since the variance of (\ref{PPP}) is the ratio $N/M$, from the distribution of the eigenvalues  of $U$  obtained numerically,  it is easy to obtain that $\alpha \simeq \pi$, thus $M \simeq k_0L=2\pi a$.

Finally, we note that this result, namely  the  spectrum of $U_{ij}=J_0(k_0r_{ij})$ is approximated by the Marchenko-Pastur law, should not come as a surprise since the matrix $T$ is being proportional to the unit matrix (\ref{Tmatrix}). Therefore, we expect from (\ref{decomposition}) that $P ( \Gamma)$ obeys the Marchenko-Pastur law.

\subsection{ B. Scaling function}

After obtaining the distribution of photon escape rates, we calculate the scaling function $C$, defined in (\ref{C}). The function $C$ has already been obtained in Section IV in the two limiting cases, namely the Dicke  and the dilute gas limits, and now we are interested in the case where $a\geq1$.

The behavior of $C$ as a function of  the system size $a$ and dimensionless density $W$ is presented in Fig. 3.
The results collapse on a single curve (Fig. 4) when plotted as a function of $N/M=aW$, indicating that the photon undergoes a crossover from delocalization towards localization as the scaling variable $N/M$  is increased. 

With the help of  the first term of (\ref{PPP}), we can approximate $C$ for $N/M\geq 2$  by
\be C\simeq1-2\frac{M}{N},\label{CC} \ee as presented in Fig. 4. 

\begin{figure}[ht]
\centerline{ \epsfxsize 9cm \epsffile{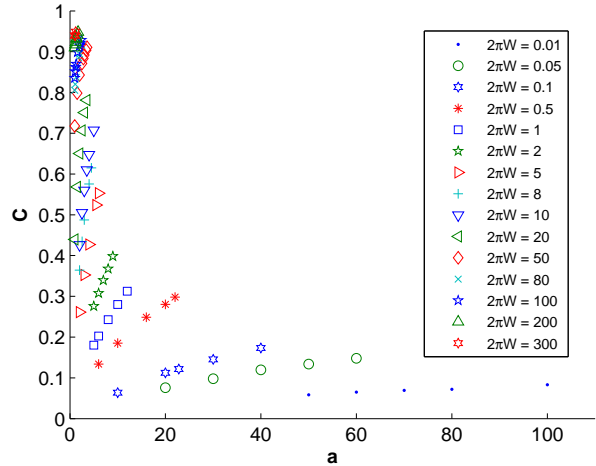} } \caption{\em (Color
   online)Behavior of the scaling function $C$ as a function  of the system size $a\geq1$ and the dimensionless density $W$.}
 \label{fig3}
\end{figure}

\begin{figure}[ht]
\centerline{ \epsfxsize 9cm \epsffile{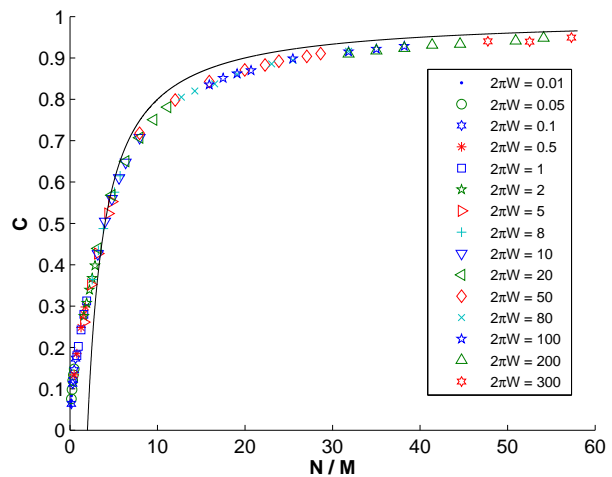} } \caption {\em (Color
   online)Behavior of the scaling function $C$ as a function of the scaling variable $N/M=aW$. All the points represented in Fig. 3 collapse on the same curve. The solid line is given by (\ref{CC}).}
 \label{fig4}
\end{figure}

\section{VI. Discussion}

The distribution  of the eigenvalues $\Gamma$  in the two-dimensional case is  similar to the one obtained in  the three-dimensional case \cite{PRL2,skipetrov}, but qualitatively different from the spectrum of the one-dimensional case \cite{EPL}. As discussed in Section III, this is due to the dependence of  $U$  on the inter-atomic distance in the dilute gas limit. Since in the two- and three-dimensional cases $U$  falls off with the square root of the inter-atomic separation and the inter-atomic separation, respectively, the single-atom limit can be reached. The one-dimensional coupling matrix, however, is a periodic function of the inter-atomic distance, hence the single-atom limit cannot be obtained. The opposite  limit, namely the Dicke limit, is reached in all dimensions.

Away from the Dicke limit, the function $C$ exhibits a scaling behavior in all dimensions, and is approximated by (\ref{CC}) in  two and three dimensions. In one dimension  the scaling function  given  by (\ref{CC}) is exact  and $M=2$ \cite{EPL}. In the two-dimensional case, considered in this paper, $M\simeq k_0L$ and in three dimensions $M \simeq (k_0L)^2$ \cite{PRL2,skipetrov}. The function $C$ in the one-dimensional case depends only on the number of atoms of the gas, unlike in  other dimensions. This dependence leads to the absence of the single-atom limit in one dimension, as for $N\gg1$, always $C=1$.
Due to the different dependence of $M$ on $k_0L$, for a given value of the latter,  localization occurs earlier in the two-dimensional case compared to the three-dimensional case. This last result stems from the different dependence of $U$  on the inter-atomic distance in the dilute gas limit, discussed earlier. 
  
In the Dicke limit, $C$ exhibits a scaling behavior as well, and is given by $C=1-2/N$ in all dimensions. Thus, for large number of atoms, the photon is always localized.

\section{VII. Conclusions and outlook}

We have studied photon emission rates from a two-dimensional atomic gas while taking into account cooperative effects between the scatterers.
To this purpose we have considered $N\gg1$ identical atoms, placed at random positions on a square in a scalar radiation field.
Based on the spectrum of the random Euclidean matrix $U_{ij}=J_0(x_{ij})$, where  $x_{ij}$ is the dimensionless random distance between any two atoms,  a scaling behavior for the escape rates is obtained. 
We have shown that away from the Dicke limit the photon undergoes a crossover from delocalization towards localization as the scaling variable $N/M$ is increased. Such behavior is  similar to the one obtained in  the three-dimensional case, but qualitatively different from the result obtained in one dimension.

The behavior obtained in  two and three dimensions provides an interesting link to the widely studied "small world networks" \cite{strogatz1,barabasi}. These networks interpolate between a regular (or ordered) lattice and a random network and combine their properties.
In a  regular lattice each vertex (out of $N$) is connected to the $K$ vertices closest to it, while in a random network the links are distributed randomly.
The small world network, tuned to be intermediate between an ordered lattice and a random network, provides a model for the topology of a wide variety of systems, such as the internet, neural networks, and coupled oscillators.

 It has been shown \cite{amaral,moore} that small world networks exhibit a crossover, rather than a phase transition, between regular lattices, characterized by long ($\propto N$) path lengths \cite{fn} and random networks with short ($\propto \ln N$) ones.
 Since small world networks appear to be  suited to  synchronize  non-linear oscillators \cite{strogatz2} and the fact that cooperative emission stems from the synchronization of the atomic dipoles induced by long range correlations, this connection becomes even more interesting.


\acknowledgments
 The authors acknowledge support by the Israel Science Foundation Grant No. 924/09.


\begin{thebibliography}{99}
\bibitem{note} The $\delta$ function is considered as a limiting case of an even function peaked at zero. Thus, the integral of $\delta(x)$ from 0 to infinity equals 1/2.
\bibitem{fn} The path length is defined as the average shortest path between any randomly chosen vertices.
\end{thebibliography}
\end{document}